\documentclass[a4paper,11pt]{article}
\usepackage{pos}

\title{Theoretical Highlights of CP Violation in $B$ Decays}
\author*[a,b]{Eleftheria Malami}

\affiliation[a]{Center for Particle Physics Siegen (CPPS), Theoretische Physik 1, Universit\"at Siegen,\\
   D-57068 Siegen, Germany}

\affiliation[b]{Nikhef,\\
Science Park 105, 1098 XG Amsterdam, Netherlands}

\emailAdd{Eleftheria.Malami@uni-siegen.de}
\emailAdd{emalami@nikhef.nl}
\abstract{In this presentation, we discuss recent key topics in theoretical analyses of CP violation in benchmarks decays of the  $B$ meson. We provide the most updated values of the mixing phases and discuss the importance of including the penguin contributions in their studies. Exploring intriguing patterns in purely tree decays, interesting new methodologies can be developed and applied. New data related to the CP asymmetries of key modes like $B^0_d \to \pi^0 K_S$ and $B^0_s \to K^+ K^-$ lead to interesting results. The new $R_{K^{(*)}}$ measurement, compatible with the Standard Model, can still allow for electron-muon symmetry violation through new sources of CP violation.}

\FullConference{
16th International Conference on Heavy Quarks and Leptons (HQL2023)\\
28 November-2 December 2023\\
TIFR, Mumbai, Maharashtra, India\\}

\begin{document}
\maketitle

\section{First Few Words}
The concept of CP violation in the $B$--meson system is important in order to test the Standard Model (SM) as well as to search for hints of New Physics (NP). Central role in the studies of CP violation plays the Cabibbo--Kobayashi--Maskawa (CKM) matrix \cite{Cabibbo:1963yz,Kobayashi:1973fv}, and an important topic linked to this matrix is the Unitarity triangle (UT). We provide recent highlights in the studies of CP violation and discuss benchmark modes.

\section{Topic 1: Mixing Phases and Penguin Contributions}
An important feature of neutral $B$ mesons is the oscillation between $B^0_q$ and $\bar B^0_q$, which may lead to interference effects if both $B^0_q$ and $\bar B^0_q$ decay into the same final state. These interference effects give rise to CP-violating asymmetries. Associated to the $B^0_q$--$\bar B^0_q$ mixing phenomenon are the CP-violating mixing phases $\phi_d$ and $\phi_s$ for the $B_d$ and $B_s$ systems, respectively. Benchmark decays for determining these phases are the $B^0_d \to J/ \psi K^0_s$ and $B^0_s \to J/ \psi \phi$ modes. The theoretical precision is limited by doubly-Cabibbo suppressed penguin contributions, which are difficult to calculate. So, instead of calculating these penguin topologies, we use control channels to determine them, exploiting the SU(3) flavour symmetry of strong interactions. The corresponding analysis is given in Ref.~\cite{Barel:2020jvf} and allows us to extract the values of the mixing phases, taking the penguin contributions into account. The experimental input for the phases is: 
\begin{equation}
    \phi_s^{\text{eff}} = (-4.1 \pm 1.3)^{\circ}, \quad \phi_d^{\text{eff}} = (43.6 \pm 1.4)^{\circ},
\end{equation}
where $ \phi_q^{\text{eff}}= \phi_q + \Delta \phi_q$ with $ \Delta \phi_q$ indicating the hadronic phase shift.
The extracted penguin parameters $a_{(V)}$ and $\theta_{(V)}$ and mixing phases are the following  \cite{Barel:2022wfr}:
\begin{align}\label{eq:Master_VP}
{\text{vector-pseudoscalar states \ \  }}    &a = 0.14_{-0.11}^{+0.17}\:, \qquad
    \theta = \left(173_{-45}^{+35}\right)^{\circ}\:, \qquad
    \phi_d = \left(44.4_{-1.5}^{+1.6}\right)^{\circ}\:,  \\
        {\text{vector-vector states \ \  }} &a_V = 0.044_{-0.038}^{+0.085}\:, \quad
    \theta_V = \left(306_{-112}^{+\phantom{0}48}\right)^{\circ}\:, \quad
    \phi_s = \left(-4.2 \pm 1.4\right)^{\circ}\:.
\end{align}
We note that updated measured values for $ \phi_s^{\text{eff}}$ and $ \phi_d^{\text{eff}}$ have recently been provided by the LHCb Collaboration and the Belle II experiment in Refs.~\cite{LHCb:2023sim,LHCb:2023zcp,Belle-II:2023nmj}, respectively. In future analyses, achieving much higher precision, it is important that the penguin contributions are properly included.

\begin{figure}[t!]
    \centering
    \includegraphics[width=0.49\textwidth]{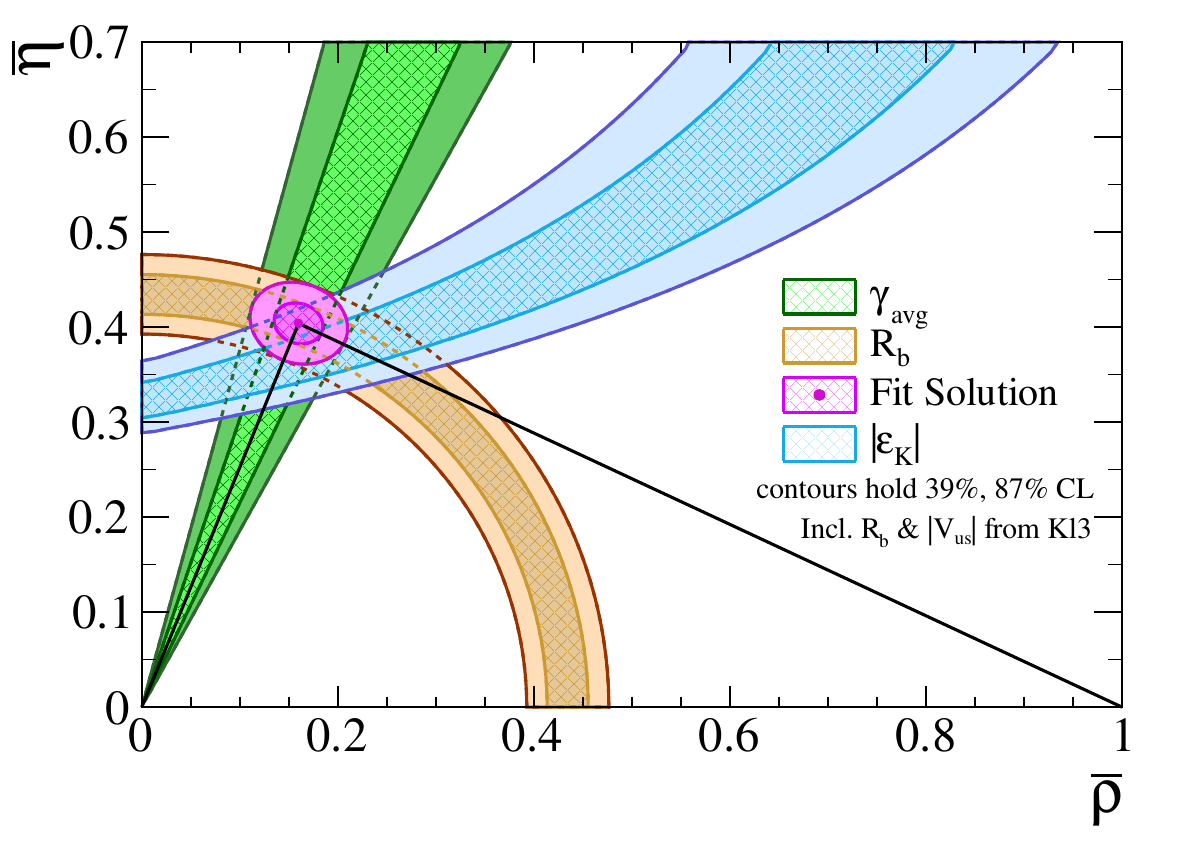}
    \includegraphics[width=0.49\textwidth]{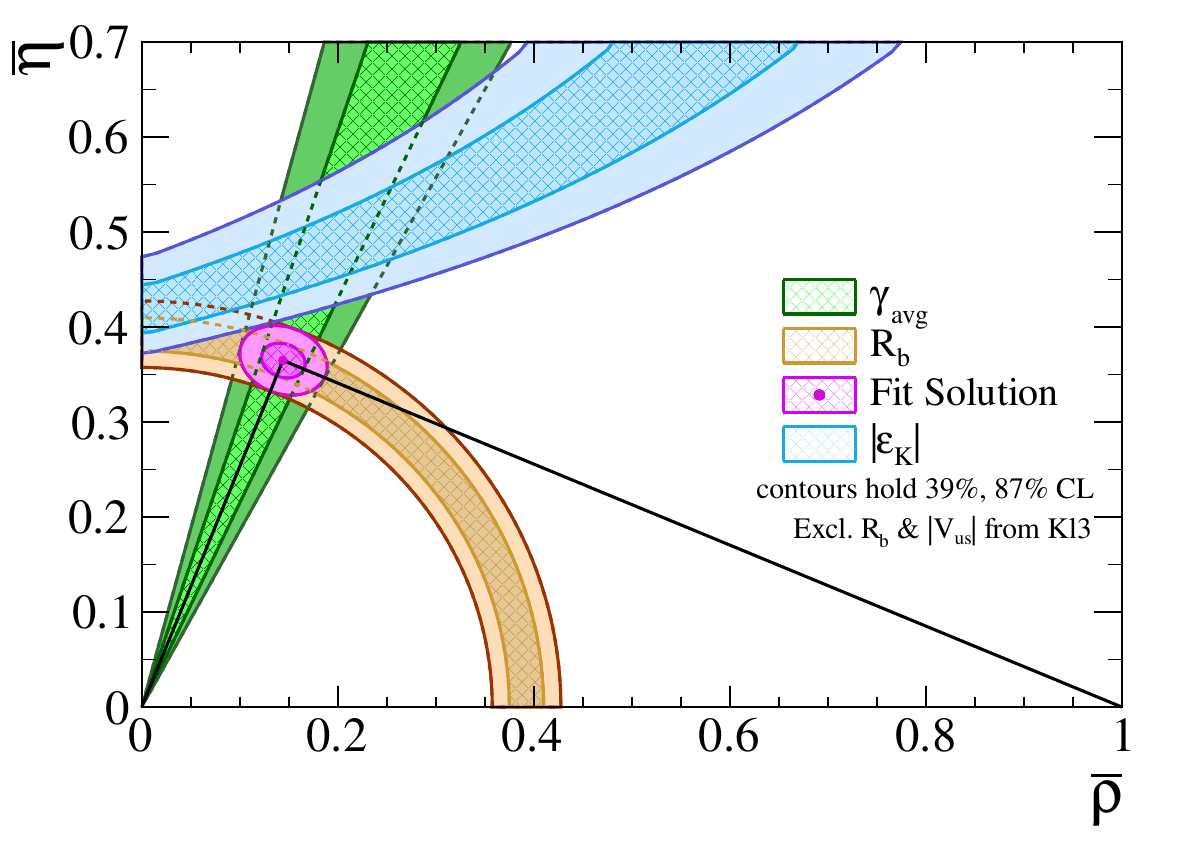}
    
    \includegraphics[width=0.49\textwidth]{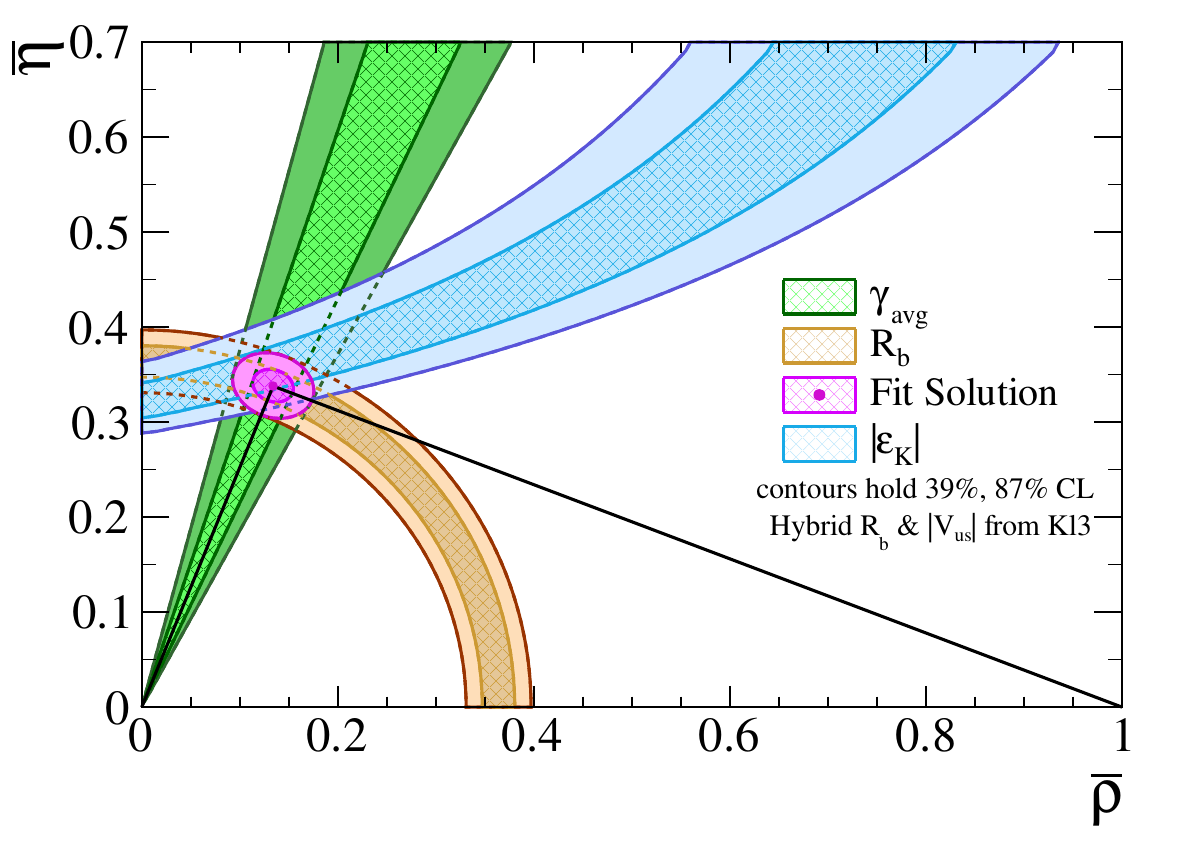}
    \caption{Determination of the UT apex. Top left: inclusive case, top right: exclusive case  and 
    bottom plot: hybrid case with excl. $|V_{ub}|$ and incl. $|V_{cb}|$ \cite{Barel:2020jvf}.
    }
    \label{fig:UT_apex}
\end{figure}

Having provided the mixing phases, we move on to the UT and the determination of its apex. For this aim, special attention needs to be given to the determination of the CKM input parameters. As presented in Ref.~\cite{DeBruyn:2022zhw}, one way of determining the UT is through the angle $\gamma$ and the side $R_b$. Concerning the angle $\gamma$, it is measured by the LHCb collaboration through $B \to DK$ modes. An alternative way of obtaining this angle is via the isospin analysis of $B \to \pi \pi,  \rho \pi, \rho \rho$ decays, yielding the UT angle $\alpha$. Utilising the $\phi_d$ phase, the value of $\alpha$ is converted into $\gamma$. The two results agree with each other, thus with the current precision, we can make an average of these $\gamma$ values:
\begin{equation}
    \gamma_{\text{avg}} = (68.4 \pm 3.3)^{\circ}.
\end{equation}
Regarding $R_b$, tensions arise between the inclusive and exclusive determinations of the $|V_{ub}|$ and $|V_{cb}|$ matrix elements. The essential point is to avoid making averages between these values but to perform separate analysis for the two different approaches. On top of that, we can explore a hybrid option; the case of exclusive $|V_{ub}|$ and inclusive $|V_{cb}|$, as presented in \cite{DeBruyn:2022zhw}. An illustration of the UT for these three cases is given in Fig.~\ref{fig:UT_apex}. In the same figure, the $\varepsilon_K$ hyperbola (blue contour), coming from indirect CP violation in the neutral kaon system, is also shown.

The studies of the UT are a key input for obtaining SM predictions of the $B_q$ mixing parameters and eventually, exploring the corresponding space for NP left through the current data. As discussed in \cite{DeBruyn:2022zhw}, the corresponding results have interesting applications in rare leptonic decays. In particular, we can minimise the impact of the CKM parameters when constraining NP in $B^0_s \to \mu^+ \mu^-$ by creating ratios of the branching fraction of this decay and the mass difference $\Delta m_s$ \cite{Buras:2003td,Buras:2021nns}:
\begin{equation}
R_{s\mu}= \bar{\mathcal{{B}}}(B^0_s \to \mu^+ \mu^-)/ \Delta m_s,
\end{equation}
where the CKM elements drop out in the SM. 

\section{Topic 2: Puzzles in Tree Decays}
The pure tree decays $\bar{B}^0_s\to D_s^+K^-$ and $B^0_s\to D_s^+ K^-$  (and their CP conjugates) offer a powerful probe for testing the SM description of CP violation \cite{ADK,RF-BsDsK,DeBFKMST}. Intriguing puzzles arise in the angle $\gamma$ of the UT and the individual branching ratios, which complement each other. A strategy that allows us to study these anomalies and search for hints of NP is discussed in Refs.~\cite{Fleischer:2021cwb,Fleischer:2021cct}. Let us summarise the key points of this methodology.

As a first step, we explore CP violation. Due to $B^0_q$--$\bar B^0_q$ mixing, interference effects arise between the $\bar{B}^0_s\to D_s^+K^-$ and $B^0_s\to D_s^+ K^-$ channels. These interference effects lead to a time-dependent CP asymmetry, which yields the observables $C$, $S$, $ \mathcal{A}_{\Delta \Gamma}$ and their CP conjugates. A measure of the strength of the interference effects is given by the quantities $\xi$ and $\bar{\xi}$. Therefore, one can use the observables $C$, $S$, $ \mathcal{A}_{\Delta \Gamma}$ and $\bar{C}$, $\bar{S}$, $ \mathcal{\bar{A}}_{\Delta \Gamma}$, in order to determine $\xi$ and $\bar{\xi}$, respectively, from the experimental data in an unambiguous way. Within the SM, in the product $\xi \times \bar{\xi}$ hadronic matrix elements cancel out allowing a theoretically clean extraction to $(\phi_s + \gamma)$. In the presence of NP, the generalisation of this relation takes the following form: 
\begin{equation}
\xi \times \bar{\xi}  = \sqrt{1-2\left[\frac{C+\bar{C}}{\left(1+C\right)\left(1+\bar{C}\right)}
\right]}e^{-i\left[2 (\phi_s +\gamma_{\rm eff})\right]},
\label{eq:generxi}
\end{equation}
where again hadronic uncertainties cancel. In particular, here it is possible that $C+\bar{C}$ is not equal to $0$, as in the SM. The above expression leads to a theoretically clean determination of the angle:
\begin{equation}
\gamma_{\rm eff}\equiv \gamma + \gamma_{\text{NP}} ,
\end{equation}
where $\gamma_{\rm NP}$ is a function of the NP parameters $\rho$, $\varphi$, ${\delta}$, and $ \bar{\rho}, \bar{\varphi}, \bar{\delta} $ (for the CP conjugate case). Here, $\rho = \left[{ A(\bar{B}^0_s \rightarrow D_s^+ K^-)_{{\text{NP}}} }/{  A(\bar{B}^0_s \rightarrow D_s^+ K^-)_{{\text{SM}}} } \right]$ measures the strength of NP, while ${\delta}$ and ${\varphi}$ denote the CP-conserving and CP-violating phases, and similarly for $ \bar{\rho}, \bar{\varphi}, \bar{\delta}$. Using information on $\gamma$ \cite{Amhis:2019ckw} from other processes, we extract $\gamma_{\rm NP}$. 

The second step corresponds to information from the branching ratios. We create ratios by combining the branching fractions of the non-leptonic decays we study with differential branching ratios of their semi-leptonic partner channels. These ratios with the semileptonic decays minimize the dependence on the CKM matrix elements and the hadronic form factors. Therefore, they provide a useful setup which permits the extraction of the colour factors $|a_1|$ from the data in the theoretically cleanest possible manner. Comparing these experimental results with theoretical predictions, we find tensions even up to the $4.8~\sigma$ level. This intriguing pattern is in line with what we expect from the puzzling situation with $\gamma$. In order to interpret these $|a_1|$ deviations, we introduce the quantities:
\begin{equation}\label{b-bar-def1}
\bar b \equiv  \frac{\langle\mathcal{B}(\bar {B}^0_s \rightarrow D_s^+K^-)_{\rm th}  \rangle}{\mathcal{B}(\bar{B}^0_s \rightarrow D_s^{+}K^{-})_{\rm th}^{\rm SM}}  = 1+2 \, \bar\rho\cos\bar\delta\cos\bar\varphi + \bar\rho^2,
\end{equation}
\begin{equation}\label{b-def}
b\equiv  \frac{\langle\mathcal{B}({B}^0_s \rightarrow D_s^-K^+)_{\rm th}  \rangle}{\mathcal{B}({B}^0_s \rightarrow D_s^{-}K^{+})_{\rm th}^{\rm SM}}  = 
1+2 \, \rho\cos\delta\cos\varphi + \rho^2,
\end{equation}  
where now we use as input the theoretical expectation of $|a_1|$. The extracted values of $b$ and $\bar{b}$ deviate from the SM. We highlight that making use of other control channels, we are able to constrain the contributions from exchange diagrams and no anomalous enhancement is observed due to these topologies.

Last but not least, we explore how much room there is for NP utilising all three $\gamma_{\rm eff}$, $b$ and $\bar{b}$. More specifically, we obtain correlations between the NP parameters $\rho(\varphi)$ and $\bar{\rho}(\bar{\varphi})$, assuming that the strong phases equal to $0$. Constraining these NP parameters, we find that it is possible to accommodate the current data with new contributions of moderate size.

We note that for the numerical analysis in Refs.~\cite{Fleischer:2021cwb,Fleischer:2021cct} the values presented by the LHCb Collaboration in Ref.~\cite{BsDsK-LHCb-CP} have been used. A new standalone measurement by LHCb using only Run II has recently been reported  \cite{LHCb:2023mcw}, which is interesting to explore further.

\section{Topic 3: $B^0_d \to \pi^0 K_S$ and $B_s \to K^+ K^-$}
Another powerful probe for CP violation studies is given by charmless two-body $B$ decays, such as $B \to \pi K$, $B_{(s)} \to KK$ and $B \to \pi \pi$ \cite{Neubert:1998pt,Buras:1998rb,Huber:2021cgk}. Here, we will present updates related to two of these channels, $B^0_d \to \pi^0 K_S$ and $B_s \to K^+ K^-$, which are dominated by penguin topologies.

The $B^0_d \to \pi^0 K_S$ decay is one of the most interesting $B \to \pi K$ channels, as this is the only one exhibiting mixing-induced CP violation. Therefore, it is important  to measure CP violation with the highest precision in this system, and especially the mixing-induced CP violation. This mode is extensively studied in Refs.~\cite{Fleischer:2018bld,Fleischer:2017vrb,Fleischer:2008wb,Buras:2004ub}.

Following the analysis in Ref.~\cite{Fleischer:2018bld}, utilising an isospin relation and complementing it with a minimal SU(3) input, we obtain correlations between the CP asymmetries, given by the following expression \cite{Faller:2008gt}:
\begin{equation}
{S_{\pi^0K_s} = \sqrt{1 - A^2_{\pi^0 K_s}} \sin(\phi_d - \phi_{00})}.
\label{S-A}
\end{equation}
Here, $S_{\pi^0K_s}$ is the mixing-induced and $A_{\pi^0 K_s}$ the direct CP asymmetry, $\phi_d$ is the $B^0_d -\bar{B}^0_d$ mixing phase and $\phi_{00}$ denotes the angle between the decay amplitude $B^0_d \to \pi^0 K^0$ and its CP-conjugate $\bar{B}^0_d \to \pi^0 \bar{K}^0$. Interestingly, tensions arise with the SM picture. Thus, we need to explore how this puzzle can be resolved. Two are the options: either the data should change or NP physics contributions might enter the penguin sector.

An update on the time-dependent CP violation in $B^0_d \to \pi^0 K_S$ was recently provided by Belle II. The new new results for the mixing induced and direct CP asymmetries are the following \cite{Veronesi:2023dak}: 
\begin{equation}
A_{\pi^0 K_s}^{\text{Belle II}}=0.04^{+0.15}_{-0.14}\pm 0.05, \quad  S_{\pi^0K_s}^{\text{Belle II}}=0.75^{+0.20}_{-0.23}\pm 0.04.
\end{equation}
These results can be compared with the current world average:
\begin{equation}
A_{\pi^0 K_s}^{\text{world average}}=-0.01\pm0.10, \quad S_{\pi^0 K_s}^{\text{world average}}=0.57\pm0.17.
\end{equation}
The new Belle II data become competitive with the world's most precise measurements. In comparison with the theoretical results for the CP asymmetries derived from Eq.~\ref{S-A}, this new measurement has been shifted towards the theory predictions, showing a better agreement within the uncertainties. This is an interesting point as it can play a key role in resolving the longstanding $B \to \pi K$ puzzle.

The second interesting channel we discuss is the $B_s \to K^+ K^-$ \cite{Fleischer:1999pa}, where the first observation of CP violation in this decay was recently reported by the LHCb collaboration \cite{LHCb:2020byh}. The new LHCb measurements reveal surprising differences between the direct CP asymmetries in the following modes \cite{LHCb:2020byh}:
\begin{align}
	\mathcal{A}_{\rm CP}^{\rm dir}(B_s^0 \to K^- K^+) - \mathcal{A}_{\rm CP}^{\rm dir}(B_d^0 \to \pi^- K^+) &= 0.089 \pm 0.031 \ , \label{eq:aCPdirDiff1}  \\
	\mathcal{A}_{\rm CP}^{\rm dir}(B_d^0 \to \pi^- \pi^+) - \mathcal{A}_{\rm CP}^{\rm dir}(B_s^0 \to K^- \pi^+) &= -0.095 \pm 0.040. \label{eq:aCPdirDiff2}
	\end{align}
These decays differ only via the spectator quark thus, it is unlikely that this pattern indicates NP. An analysis performed in Ref.~\cite{Fleischer:2022rkm} shows that these differences can be accommodated in the SM through exchange and penguin-annihilation topologies, that are sizeable $-$ at the level of $20\%$. 
 
 On top of that, the analysis in Ref.~\cite{Fleischer:2022rkm} provides an interesting way of extracting the angle $\gamma$ of the UT. The proposed strategy relies only on CP asymmetries without requiring information on branching ratios. The result:
 \begin{equation}
 \gamma = (65^{+11}_{-7})^\circ,
 \end{equation}
agrees excellently with the $\gamma$ values from the $B\to D K$ decays, which are pure tree transitions. As an a alternative, the  $B^0_s$--$\bar B^0_s$ mixing phase $\phi_s$ can also be determined if now one uses the value of $\gamma$ as an input. For this purpose, the methodology of using ratios of non-leptonic and semileptonic $B_{(s)}$ decay rates is utilisied, providing a clean way of obtaining $\phi_s$.
 
\section{Topic 4: CP Violation in Rare Decays. What about $R_{K^{(*)}}$ and the Electron-Muon Symmetry Violation?}
The new results for $R_{K^{(*)}}$ presented by the LHCb collaboration in 2022 \cite{LHCb:2022qnv,LHCb:2022vje}:
\begin{equation}\label{eq:rkmeas}
    \langle R_K \rangle = 0.949\pm 0.05, {\text{ \ for momentum transfer }} q^2 \in [1.1, 6.0]\  {\text{GeV}},
   \end{equation}
brought new perspectives for testing the electron--muon universality. These results agree with Lepton Flavour Univerasality (LFU). The differential rates for $B\to K \mu^+\mu^-$ are small compared to the SM predictions, deviating at the $3.5 \, \sigma$ level, thereby still indicating possible NP through these decays. How much electron-muon universality violation is possibly left for this NP, now constrained by $R_K$? As shown in the analysis in Ref.~\cite{Fleischer:2023zeo} due to new CP-violating effects, there is still significant room for violation of the electron--muon universality. 

More specifically, this analysis explores the CP-violating effects in the NP studies of rare decays making use of the muonic Wilson coefficients. The experimental branching ratio and the direct CP asymmetry of the $B^ -\to K^- \mu^+\mu^-$ mode constrains the corresponding muonic Wilson coefficients $C_{i \mu}$. Combining this constrain with the new $\langle R_K \rangle$ measurement allows the determination of the Wilson coefficient $C_{i e}$ in the electronic sector. Having determined $C_{i e}$, the electronic direct and mixing-induced CP asymmetry are also obtained. The following conclusions are drawn:
\begin{itemize}
\item [i)] NP Wilson coefficients entering the electronic modes can strongly differ from the corresponding ones entering the muonic channels and 
\item [ii)] CP violating phases can be significantly different, therefore also the CP asymmetries between the electronic and muonic modes, which are the observables that they experimentalists should utilize in order to test the violation of LFU.
\end{itemize}

Therefore, it is still possible to have significant electron--muon universality violation, if NP effects are associated with new sources of CP violation, which are encoded in Wilson coefficients. In the era of high-precision B physics, it is important to perform experimental searches focusing on differences in CP asymmetries between the $b \to s e^+ e^-$ and $b \to s \mu^+ \mu^-$ transitions. These studies will be essential for further testing LFU.

\section{Epilogue}
A lot of progress has been achieved over the last years in the studies of CP violation, which was possible through the synergy between theorists and experimentalists. There are exciting new perspectives to further explore CP violation. Moving towards the high precision era of $B$ physics and monitoring the evolution of the data will lead to a much sharper picture. 

\section*{Acknowledgements}
\noindent I would like to thank the organisers of the HQL2023 for the invitation and the opportunity to attend such a great conference. %Special thanks to our colleague Kristof de Bruyn, who has updated the numerics regarding the mixing phases and the UT plots -derived using the tool GammaCombo- following the analysis of the project we had with Robert Fleischer and Philine van Vliet.

\end{document}